\documentclass[useAMS,usenatbib]{mn2e}

\usepackage{amsmath}
\usepackage{amssymb}
\usepackage[dvips]{graphicx}
\usepackage{epsfig}

\begin{document}
\title {An estimate of the temporal fraction of cloud cover at  San Pedro M\'artir Observatory}

\author[E. Carrasco et al.]
{E. Carrasco$^{1}$,\thanks{E-mail: bec@inaoep.mx}
A. Carrami{\~n}ana$^{1}$,
L. J. S{\'a}nchez$^{2}$,
R. Avila$^{3,4}$ and
I. Cruz-Gonz{\'a}lez$^{2}$\\
$^{1}$ Instituto Nacional de Astrof{\'\i}sica, \'Optica y Electr\'onica,
Luis Enrique Erro 1, Tonantzintla, Puebla 72840,  M\'exico\\
$^{2}$  Instituto de Astronom\'{\i}a, Universidad Nacional Aut\'onoma de M\'exico, Apartado Postal
70--264, Cd. Universitaria 04510, M\'exico D.F., M\'exico\\
$^3$ Centro de F{\'\i}sica Aplicada y Tecnolog{\'\i}a Avanzada, Universidad Nacional
 Aut\'{o}noma de M\'{e}xico,  A.P. 1-1010,\\ Santiago de Quer\'{e}taro, Qro. 76000\\
$^4$  Centro de Radioastronom\'{\i}õa y Astrof\'{\i}sica, Universidad Nacional
Aut\'{o}noma de M\'{e}xico,\\ Apartado Postal 3-72, Morelia, Michoac\'an 58089, M\'exico}
\date{Accepted 2011 November}

\pagerange{\pageref{firstpage}--\pageref{lastpage}}
\pubyear{2012}

\maketitle

\label{firstpage}

\begin{abstract}
San Pedro M\'artir in the Northwest of Mexico is the  site of the Observatorio Astron\'omico
Nacional. It was one of the five candidates sites for the Thirty Meter Telescope, whose
site-testing team spent four years measuring the atmospheric properties on site with a very
complete array of instrumentation. Using the public database created by this team, we
apply  a novel  method  to solar radiation data to estimate  the   daytime fraction of
time when the sky is clear of clouds. We analyse  the diurnal, seasonal  and annual cycles of
cloud cover. We find that  82.4 per cent of the time the sky is clear of clouds. Our
results are consistent
with those obtained by other authors, using different methods, adding support to this value and proving
the potential of the applied method. The clear conditions at the site are particularly good
showing that San Pedro M\'artir is an excellent  site  for optical and infrared  observations.
\end{abstract}


\section{Introduction}
 The San Pedro M\'artir (SPM)  observatory is located at
$31^{\circ} 02^{\prime} 39^{\prime \prime}$N,
$115^{\circ} 27^{\prime} 49^{\prime \prime}$W and at an altitude of 2830~m, inside the
Parque  Nacional Sierra de San Pedro M\'artir. SPM is $\sim$65~km E of the Pacific Coast and
$\sim$55~km W to the Gulf of California.
The largest telescope at the site is a 2.1-m   Ritchey-Chr\'etien, operational since 1981.
Astroclimatological characterization studies at SPM are reviewed in~\citet{Tapia07}.
Compilations of one and two continuous decades of weather and observing statistics of OAN-SPM
have been reported by ~\citet{Tapia92} and ~\citet{Tapia03}. The yearly fractions of photometric
and spectroscopic nights from 1984 to 2006 is presented on Table 3 of \citet{Tapia07}.  Other
aspects of the site characterization have been reported  by several authors e.g.~\citet{CruzGlez03}
and~\citet{CruzGlez07}. Nevertheless,  this is the first study on the radiation data measured in situ.
The data were recorded by the Thirty Meter Telescope (TMT) site-testing team from 2004 to 2008;
see~\citet{Schock09} for an overview  of the TMT project and  its main results.

Cloud cover is one of the most important considerations to characterize a ground-based
astronomical observatory. Only for low-frequency radio observations cloudiness is of little
importance. Given a site, statistics of daytime cloud cover are indicative of the usable
portion of the time for optical and near-infrared observations and bring key information for
the potentiality of that site for millimeter and sub-millimeter astronomy. The relationship
between diurnal and nocturnal cloudiness is strongly dependent on the location of the site.
\citet{Erasmus02} give a detailed discussion on that topic. For the case of SPM these authors
conclude that the day versus night variation of the cloud cover is less than 5 per cent.
Therefore, daytime cloud cover statistics at SPM is a useful indicator of nighttime cloud
conditions.
Here we present a study of the cloud cover over SPM using an approach recently introduced
by \citet{Carrasco09}. It consists of the computation of histograms of solar radiation values
measured at the site and corrected for the zenital angle of the Sun. The data coverage is
presented in section \S2, the method is described in \S3,  the data obtained in SMP \S4, the discussion of
the results in \S5, the statistics of cloud cover in \S6 and the summary and conclusions in \S7.

\begin{figure}
\includegraphics[width=\columnwidth]{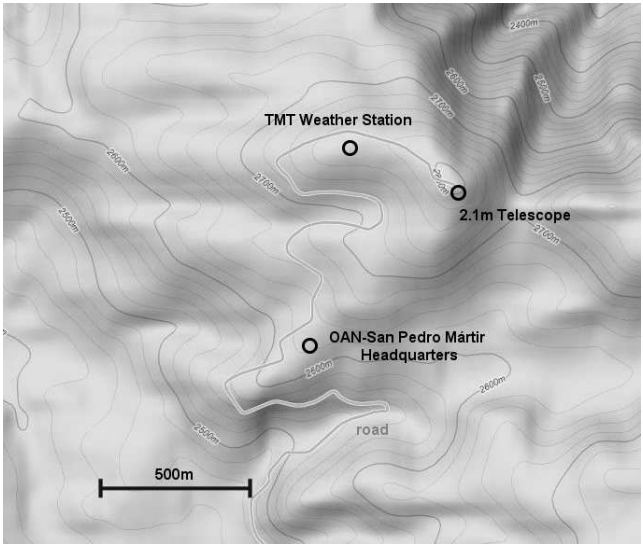}
\caption{Location of the TMT automatic weather station relative to the
the 2m telescope and the observatory headquarters. The 2.1m telescope is
at 31.04417$^\circ$N, 115.463611$^{\circ}$W.\label{location}}
\end{figure}

\section{Data coverage}
The data presented here consist of records of  solar radiation  energy fluxes in units
of Wm$^{-2}$ acquired  with an  Monitor automatic weather station  \citep{Schock09},
located 2m above the ground \citep{Skidmore09} at 31$^{\circ}$ 02$^{'}$ 43.84$^{"}$ N,
115$^{\circ}$ 28$^{'}$ 10.45$^{"}$ W.
It was in a tree free zone without visibility limitations. The location of the weather
station relative to the observatory headquarters is  shown in  Fig. \ref {location}.
The Monitor sensors have a spectral response between 400  and 950~nm with an
accuracy of 5 per cent according to the vendor. The data span between October 2,
2004 and  August 8,  2008 with a sampling time of 2 minutes.

\begin{figure}
\includegraphics[width=\columnwidth]{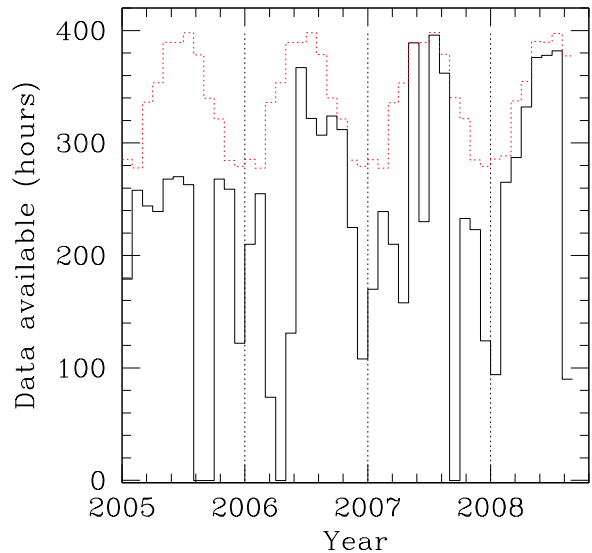}
\caption{Data coverage  in hours per month. The  red dotted lines are the hours  of
solar radiation at the site per month for the same period. \label{coverage_plus}}
\end{figure}

The analysis does not include the data from 2004, when the radiation sensor was not
working at the beginning of each day. Table \ref{coverage_per_month} summarises the
temporal coverage of the data expressed in percentage, with due considerations of the
diurnal cycle variation. The data span from  2005 January 12  to 2008 August 8, with
a 67 per cent effective coverage of the 3.6 year sample; data exist for 973 out of 1316 days.
The complete sample contains 596580 min out of 899520 possible; coverage was
59 per cent for 2005  and increased to  78 per cent  towards the end of the campaign, in 2008.
In Fig.~\ref{coverage_plus} the solid line shows  the total number of hours  per month of solar
radiation measured in situ while the red dotted line represents the corresponding
expected  radiation flux.   The coverage per  hour of day, shown in
Table~\ref{coverage_per_hour}, was obtained by  counting  the minutes of data and
comparing  with the total expected,  given SPM coordinates. Hours with very low data coverage,
less than 30 min, are  excluded in this analysis.

\begin{table}
\begin{center}
\caption{Solar radiation data coverage in percentage.\label{coverage_per_month}}
\begin{tabular}{|l|r r r r |r|}
\hline
Month       & 2005 & 2006 & 2007 & 2008  & All\\
 \hline
January    &  63  &  74  &  60  &  33   & 57\\
February   &  93  &  92  &  86  &  92   & 91 \\
March      &  73  &  22  &  63  &  85   & 61 \\
April      &  68  &   0  &  45  &  94   & 52 \\
May        &  69  &  34  & 100  &  96   & 75 \\
June       &  69  &  94  &  59  &  97   & 80 \\
July       &  66  &  81  &  99  &  96   & 86\\
August     &   0  &  81  &  96  &  24   & 50\\
September  &   0  &  95  &   0  &   -   & 32 \\
October    &  83  &  97  &  72  &   -   & 84\\
November   &  91  &  79  &  78  &   -   & 83 \\
December   &  44  &  39  &  44  &   -   & 42\\
&&&&&\\
Year Total  &  59  &  65  &  68  &  78   & {\bf 67} \\
\hline
\end{tabular}
\end{center}
\end{table}

\begin{table*}
\centering
\caption{Hourly solar radiation data coverage in percentage.\label{coverage_per_hour}}
\begin{tabular}{lrrrrrrrrrrrrr}
\hline
{} & {Jan} &  {Feb} &  {Mar} &  {Apr} &  {May} &  {Jun} &
         {Jul} &  {Aug} & {Sep} &  {Oct} &  {Nov} &  {Dec} & {Per hr} \\
\hline
  6~h    &  -   &   -  &  -   &  25  &  77  &  75  & 100  &   0  &   0  &  -   &   -  &   -   & 69\\
  7~h    &  -   &   -  &  49  &  37  &  56  &  62  &  74  &  52  &  35  &  99  &   -  &   -   & 54 \\
  8~h    &  80  & 100  &  51  &  37  &  56  &  62  &  75  &  52  &  32  &  82  &  88  &   -   & 68 \\
  9~h    &  62  & 100  &  52  &  37  &  56  &  61  &  75  &  52  &  32  &  84  &  92  &  57   & 63 \\
 10~h    &  61  &  99  &  70  &  58  &  80  &  86  &  91  &  52  &  32  &  84  &  90  &  43   & 71 \\
 11~h    &  49  &  94  &  71  &  59  &  81  &  86  &  91  &  52  &  32  &  78  &  57  &  26   & 66 \\
 12~h    &  48  &  97  &  71  &  59  &  80  &  88  &  91  &  50  &  32  &  85  &  90  &  30   & 69 \\
 13~h    &  61  &  97  &  68  &  58  &  80  &  88  &  91  &  50  &  32  &  83  &  91  &  46   & 71\\
 14~h    &  60  &  94  &  66  &  58  &  80  &  88  &  90  &  51  &  32  &  83  &  92  &  46   & 71\\
 15~h    &  60  &  93  &  65  &  58  &  80  &  88  &  91  &  51  &  32  &  81  &  93  &  46   & 70\\
 16~h    &  50  &  87  &  65  &  58  &  78  &  88  &  90  &  51  &  29  &  82  &  92  &  47   & 68\\
 17~h    &  30  &  47  &  46  &  56  &  79  &  91  &  89  &  51  &  31  &  97  &   -  &   -   & 57\\
 18~h    &   -  &  -   &   -  &  57  &  93  &  88  &  90  &  55  &  24  &  -   &   -  &   -   & 71\\
&&&&&&&&&&&&&\\
Per month &  57 &  91  &  61  &  52  &  75  &  81  &  87  &  50  &  32  &  84  &  83  &  42   & {\bf 67}\\
\hline
\end{tabular}
\end{table*}

\section{Solar radiation and inferred cloud coverage \label{radiacion}}

In this section we present  the model used to retrieve the cloud coverage from the
radiation data developed by Carrasco et al. (2009) for Sierra Negra, the site of the
Large Millimeter Telescope (LMT) and the High Altitude Water \v{C}erenkov $\gamma$-ray
observatory (HAWC).

The radiation flux at ground level is considered,  to first approximation, to be
given  by the solar constant $F_{\sun}$, modulated by the zenith angle of the Sun
and a time variable attenuation factor $\psi(t)$.
For working purposes we take $F_{\sun}=1367\,\rm Wm^{-2}$ exactly.
The solar constant varies only $\pm  1$ Wm$^{-2}$ over the 11 yr solar cycle,
\citet{Frohlich98}. These variations are negligible for the purpose of this work.
Knowing the position of the Sun  at the site as a function of time, we can estimate
the variable $\psi$, given as
\begin{equation}
\psi(t) = {F(t) \over F_{\sun}\cos\theta_{\sun}} \, .
\end{equation}
where $F(t)$ is the radiation measured at the site and $\theta_{\sun}$ is the zenith angle
of the Sun. $\psi(t)$ is a time variable factor, nominally below unity, which accounts for
the instrumental response (presumed constant), the atmospheric extinction on site and the
effects of the cloud coverage. Knowing the site coordinates, we compute the modulation
factor $\cos\theta_{\sun}$ as a function of day and local time, minute
per minute  to study  the behavior of the variable $\psi$.
In the case of Sierra Negra the histogram of values of $\psi$ showed a  clear bimodal
distribution composed by a broad component for low values of $\psi$ and a narrow peak
$\psi\la 1$. The histogram of $\psi$ values  (Fig. 11 of Carrasco et al., 2009) can be
well reproduced with a two component fit, with the first component having its maximum
around $\psi\sim 0.2$ and the narrow peak at $\psi\sim 0.75$, with the minimum at
 $\psi_{min}=0.55$ separating both components. The narrow component is interpreted as
due to direct sunshine while the broad component is originated  when solar radiation
is partially  absorbed by clouds;  we then use the relative ratio of these two
components to quantify the ``clear weather fraction".

The functional form of the fit to the histogram  of $\psi$  is given by,
\begin{equation}
f(\psi)=A\psi^{2}e^{-\beta \psi} + {B\over 1+\left[(\psi-\psi_{0})/\Delta \psi\right]^{2}}.
\label{fit}
\end{equation}
 The first term on the right hand side is a $\chi^{2}$ function with six degrees of
freedom. It is interpreted as the ``cloud-cover'' part of the data, with its
integral being the fraction of ``cloud-covered''  time. The second term,  a Lorentzian
function with centre $\psi_{0}$ and width $\Delta \psi$, represents the
``cloud-clear'' part of the data. $A$ and $B$ provide the normalization and
relative weights of both components;
$\beta$ is related to the width and centre of the broad peak.

\section{San Pedro M\'artir data}

Daily minute per minute values of the atmosphere free modulated solar radiation flux,
$F_{\sun}\cos\theta_{\sun}$, were compared with the radiation measured at the site.
Local transit cosine values range just below 1 around June 16 and July 1 (for 2005)
to 0.58 at winter solstice (December 21).

\begin{figure}
\includegraphics[width=\columnwidth]{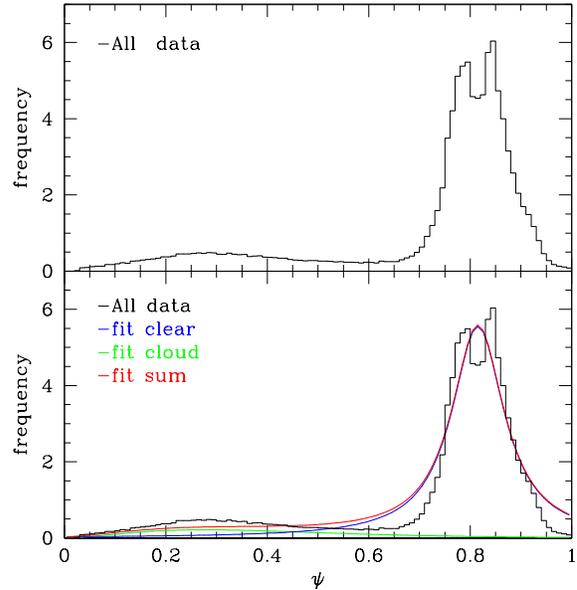}
\caption{The normalized observed distribution of $\psi$: the solar flux $F(t)$ divided
by the nominal solar flux at the top of the atmosphere, $F_{\sun}\cos\theta_{\sun}(t)$
for airmasses below 2.
{\em Top:} the distribution for all the data.
{\em Bottom:}  the fit to the data is the sum of two components: the blue line
corresponds to clear weather; the green one to cloudy weather and the red line
to the sum. The relative area of both components determines the clear/cloud fraction.
See the electronic edition of MNRAS  for a color version of this figure.
\label{rad_histfit}}
\end{figure}

\begin{figure}
\includegraphics[width=\columnwidth]{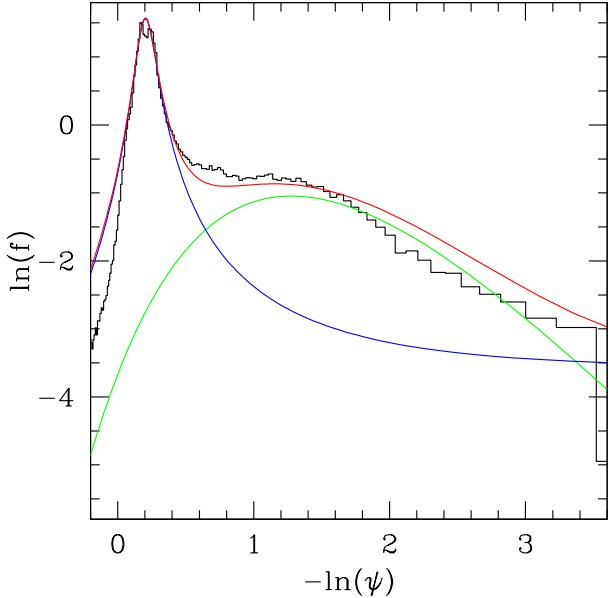}
\caption{The logarithm plot of the normalized observed distribution of $\psi$ shown in Fig.~\ref{rad_histfit}.
The blue line  corresponds to clear weather; the green one to cloudy weather and the red line
to the sum.
See the electronic edition of MNRAS  for a color version of this figure.
\label{todoslnln_psi}}
\end{figure}

For the analysis described in this section we considered  data with airmass below 2
as most  astronomical observations are carry out  at this airmass interval.
The normalized histogram of $\psi$ for all the data  is shown in the
upper panel of Fig.~\ref{rad_histfit}. It shows a double peak in the clear component,
not fully consistent with the standard narrow component fit function, and the
cloud component with maximum at $\psi \la 0.3$. We applied the double component fit of
Eq.~\ref{fit}, shown in the lower panel of Fig.~\ref{rad_histfit}. The fit for the clear
component is drawn in blue, for the cloud component in green and for the sum in red.
 The coefficients of the fit given by Eq.~\ref{fit} are presented in
Table~\ref{coefficients_global}. Fit errors were obtained through a bootstrap analysis
using 10000 samples.  The fit  agrees with the data within the statistics, except in
the wings of the clear peak in Fig.~\ref{rad_histfit}.  Still, the
Lorentzian function proved to fit much better the data than a Gaussian. The fit can be better
appreciated in a logarithm  version of Fig.~\ref{rad_histfit},  shown in  Fig.~\ref{todoslnln_psi}.

\begin{table}
\caption{Parameters of the fit shown  in red Fig.~\ref{rad_histfit}.\label{coefficients_global}}
\begin{tabular}{lccc c}
\hline
Parameter &  Global  &  Bootstrap  &  errors  & relative \\
          &   $z< 2$ &               &         & error $(10^{-3}$)  \\
\hline
$A$       &   40.8          &  40.766  &     $\pm 0.612$ & 15.0  \\
$\beta$   &   7.19          &  7.175   &     $\pm 0.044$ & 6.2\\
 $B$      &  6.03           &   6.035  &     $\pm 0.055$ & 9.1\\
$\psi_{0}$   &  0.815          &   0.8151 &     $\pm 0.0002$& 0.3 \\
$\Delta \psi$ & 0.063          &   0.0629 &     $\pm 0.0005$& 7.5  \\
f(clear) &   0.824          &   0.8238 &     $\pm 0.0009$ & 1.1\\
f(cloud) &   0.176          &   0.1762 &     $\pm 0.0009$ & 5.0\\
\hline \end{tabular}\end{table}

We  considered data  with $\psi\leq \psi_{min}$, where $\psi_{min}=0.58$,
as cloudy weather and data with $\psi>\psi_{min}$ as clear weather.
 The value $\psi_{min}=0.58$ corresponds to the intersection of the two
components of the function fitted
to the distribution of all data points. We computed the
fraction of clear time f(clear), as clear/(clear+cloudy).
From the global histogram we obtained a clear
fraction  for the site of   82.4 per cent.  The errors in the determination
of f(clear) and f(cloud), were also obtained by generating 10000 bootstrap samples;
they are shown in Table \ref{coefficients_global}.

\begin{figure}
\includegraphics[width=\columnwidth]{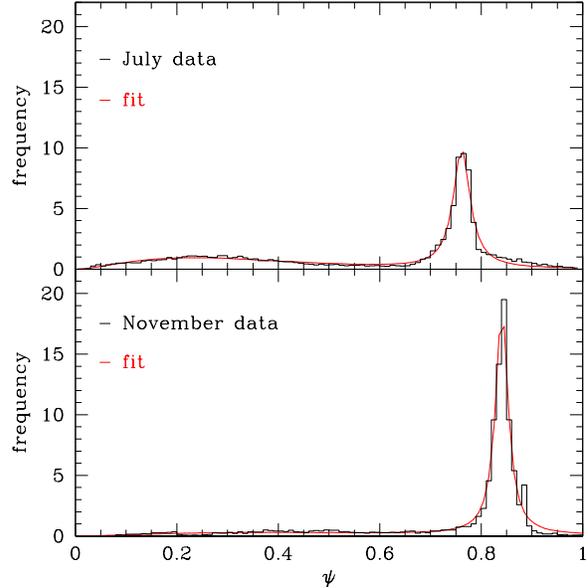}
 \caption{The observed distribution of $\psi$ and the two component fit
for July {(\em top)}  and November {(\em bottom)}. Comparing both plots a shift
in the centre of the narrow component can be appreciated. $\psi$.
See the electronic edition of MNRAS for a color version of this figure. \label{rad_histfit_jn}}
\end{figure}

\begin{figure}
\includegraphics[width=\columnwidth]{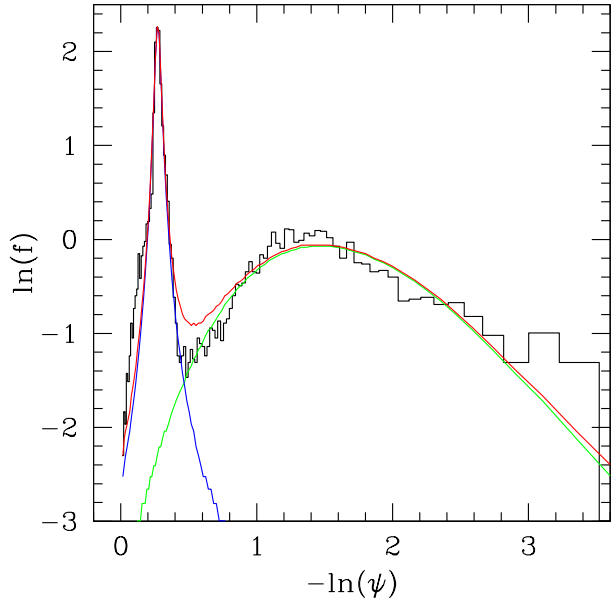}
\caption{The logarithm plot of the  observed distribution of $\psi$  for July shown in
the top panel of Fig.~\ref{rad_histfit_jn}. The fit to the data is the sum of two components:the blue line
corresponds to  clear weather; the green one to  cloudy weather and the red line
to the sum.  See the electronic edition of MNRAS  for a color version of this figure.
\label{juliolnln_psi}}
\end{figure}

\begin{figure}
\includegraphics[width=\columnwidth]{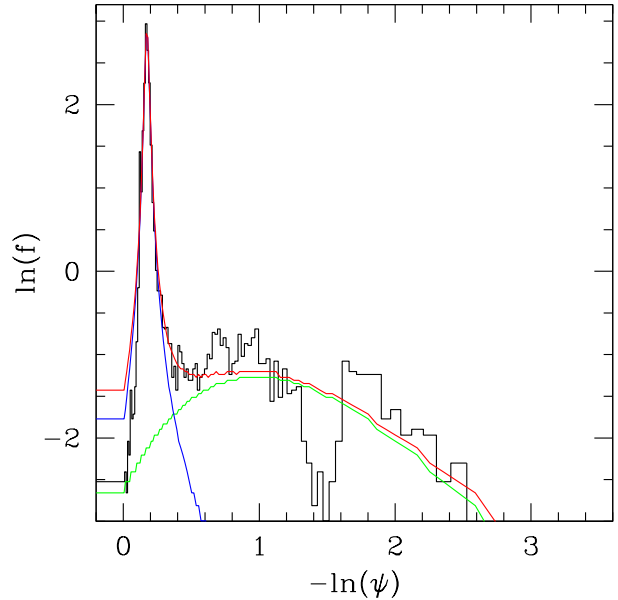}
\caption{The logarithm plot of the  observed distribution of $\psi$  for November shown in
the bottom panel of Fig.~\ref{rad_histfit_jn}. The fit to the data is the sum of two components:the blue line
corresponds to  clear weather; the green one to  cloudy weather and the red line
to the sum. Note that even in the case of low statistics  the fit proves to be good.
See the electronic edition of MNRAS  for a color version of this figure.
\label{novlnln_psi}}
\end{figure}

 To study the seasonal variation of $\psi$ we created histograms and the corresponding fits  per month. Consider
the histogram  and corresponding  fit for July and  November shown Fig.~\ref{rad_histfit_jn}.
It can be appreciated that the fits reproduce the distribution of $\psi$ very well.
The narrow clear component is consistent with prevailing
clear sky conditions, for which the
solar radiation reaches the site with only the attenuation of the atmosphere.  The coefficients
of the fits presented in Fig. \ref{rad_histfit_jn}, according to
the functional form of $f(\psi)$ given by  Eq.~\ref{fit}, are shown in Table~\ref{coefficients}.
 The fits can be  better valued in the logarithm displays of Fig.~\ref{rad_histfit_jn}, presented in
Figs.~\ref{juliolnln_psi} and \ref{novlnln_psi}. The fits  to the complete data (red line), to  clear weather
(blue line) and  to cloudy weather (green line), have been included.

\begin{figure}
\includegraphics[width=\columnwidth]{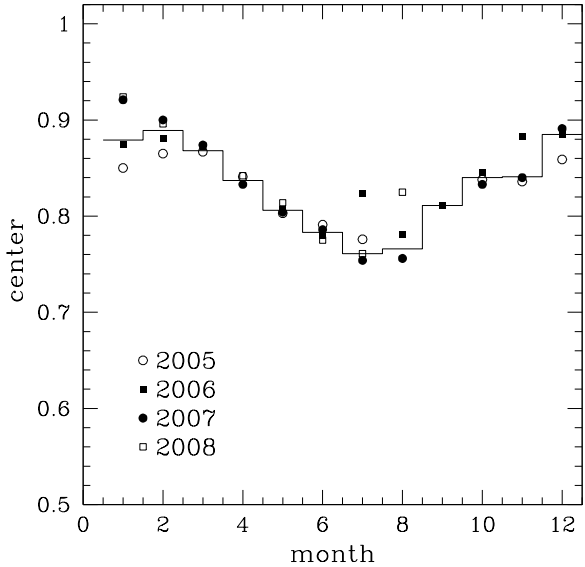}
\caption{The centre of the narrow component of $\psi$ for each month. The monthly
values for all the data are indicated by the histogram, while the dots mark
individual months of different years.
The position of the narrow peak component is not constant during the year,
reaching a minimum in July. The error bars are smaller than the symbols. \label{centers}}
\end{figure}

\begin{table}
\centering
\caption{Coefficients of the fits shown in Fig. \ref{rad_histfit_jn}.\label{coefficients}}
\begin{tabular}{l r r r r r}
\hline
  Sample  &   A   & $\beta$   &  B     & $\psi_{0}$ & $\Delta \psi $\\
\hline
  July          & 135.2   &  8.68    &  9.24  & 0.761   & 0.021 \\
                & $\pm1.5$& $\pm0.112$ &  $\pm0.141$ & $\pm0.0005$ & $\pm0.001$       \\
  November      &  11.7   &  3.83    & 19.07  &  0.841  & 0.015 \\
                & $\pm2.6$ &  $\pm0.189$ & $\pm0.236$ & $\pm0.0009$  & $\pm0.002$ \\
\hline
\end{tabular}
\end{table}

When analysing the fits per month we realised that the Lorentzian fits for the
"clear weather peak" are better than that of the complete dataset. Futhermore, we noted
a significant shift in the position of its centre, that can be clearly appreciated in
Fig.~\ref{rad_histfit_jn} where we present the plots for July and November. This effect is
also apparent in Figs.~\ref{juliolnln_psi} and \ref{novlnln_psi}.

 We studied the
position of the centre  of the peak corresponding to the clear fraction as a function of the
month of each observed year. We found that for every year there is a cyclic  effect: the centre
of the peak is  0.880  in January, reaches a maximum around 0.889 in February,
decreases to a minimum of   0.761 in July and increases  towards the end of the year
to  0.885 in December.  Errors in the statistical determination of $\psi_0$ are
$\la 0.001$. Fig.~\ref{centers} shows the position of the centre of the narrow component obtained using the
whole data set for each month   with a solid line. The corresponding values of the individual
months for different years are indicated by distinct symbols.

 The solar radiation sensor accuracy is $\pm 5\%$. However, considering  N ($\sim$20000)
data points per bin, the  position of the peaks are statistical variables determined with an accuracy
$ \propto 1/\sqrt N$ times the individual measurement error i.e. much better than  5\%.
Hence,  the variations in the position of the centres observed with an amplitude of up to 14\%
are statistically robust. Still, the  amount of radiation corresponding to the clear peak in
July is higher by 277  Wm$^{-2}$  than that received in November.

\section{Discussion}

\subsection{Global clear time  determination}

The  82.4 per cent of clear fraction obtained from the global distribution shown
 is Fig.~\ref{rad_histfit} is similar to that  reported by~\citet{Erasmus02}. In a
comprehensive study for the California Extremely  Large Telescope (CELT) project,
the authors surveyed cloud cover and water vapor conditions for different  sites
using  observations from the International Satellite Cloud
Climatology Project.  The study covers 58 months between  July 1993 and
December 1999 using a methodology that had been  tested and successfully  applied in
previous studies. They estimated that the photometric and usable fractions of
nighttime at SPM are 74  and 81 per cent, respectively. Their definition of usable
time includes conditions with high cirrus. The authors give a detailed discussion on
the relationship between diurnal and nocturnal cloudiness. For the case of SPM, they
 conclude that the day versus night variation of cloud cover
is less than 5 per cent,  being clearer  at night.

Another estimation of the useful observing time at SPM is given by ~\citet{Tapia03}
who reports a 20~yr  statistics of the fractional number of nights with totally clear,
partially clear and mostly cloudy based in the observing log file of the 2.1m telescope
night assistants. The author reports a total fraction of useful observing time of  80.8 per cent
and compares his results with those from ~\citet{Erasmus02}; he concludes that the monthly
results from both studies agree within 5 per cent while for the yearly fraction, the
 discrepancies are lower than 2.5 per cent.

\subsubsection {The diurnal cycle}

~\citet{Erasmus02} studied the diurnal cycle by calculating the average of clear time for
two sets of hours: from 8-12 (D1) and from 12-16 (D2), considering data at airmass less than
2 and defining the seasons as follows: winter: December, January and February;
spring: March, April and May;  summer: June, July and August and autumn: September, October
and November.  Our results for D1 and D2, using the same parameters,  and  theirs are shown
in Table~\ref{diurnal}.   As we are not comparing simultaneous observations
we do not expect  to obtain  necessarily  the same values of clear time for D1 \&  D2.
Nevertheless,  we  reproduce the differences between them within a few percentage, the
largest difference being  7 per cent for winter. The trend in our results for daytime is consistent
with that obtained by ~\citet{Erasmus02}.

The differences in the values for D1 and D2 and those obtained  by ~\citet{Erasmus02} are
between  3  and 28 per cent.  As a reference, in the case of the PWV, the results for the TMT
preliminary studies carried out by ~\citet{Erasmus02} and those obtained from the in situ site
testing  group are
within 30 per cent for all the sites, see \citet{Otarola10}.  Our results for D1 \& D2 are within that
range. Futhermore, the differences might be influenced  by the distinct data coverage and
by the definitions of clear time used by  ~\citet{Erasmus02}.  The results presented in this
paper are based on direct solar radiation  measurements.

\subsection{ The seasonal variation of $\psi_0$}

\begin{table}
\caption{Percentage of time that sky conditions are clear for two different periods of day
D1 \& D2.\label{diurnal}}
 \begin{tabular}{|l|r r r r r r}
 \hline
  f(clear)        &    D1   &  D2    &   D1-D2   &    D1  & D2  &  D1-D2  \\
           & \multicolumn{3}{l}{This paper}   & \multicolumn{3}{l}{E \& VS}  \\
 \hline
Summer   &       78  &   66 &    {\bf 12}    &    75  &  60  &   {\bf 15} \\
Autumn   &       88  &   84 &    {\bf 4}     &    70  &  66  &   {\bf 4} \\
Winter   &       80  &   83 &    {\bf-3}     &    52  &  62  &   {\bf -10}  \\
Spring   &       95  &   94 &    {\bf 1}     &    77  &  75  &   {\bf 2}  \\
\hline
\end{tabular}
\end{table}

The variation trend in the centre of the clear peak shown in Fig.~\ref{centers} can be
interpreted in terms of seasonal  variations of the atmospheric transmission: during the
summer months there is more   atmospheric absorption than in the rest of the year. This is
consistent with the seasonal  variation of the Precipitable Water Vapor (PWV) at
210 GHz reported by ~\citet{Hiriart97}, ~\citet{Hiriart03}, ~\citet{Otarola09,Otarola10}.
The seasonal variation in the
PWV is shown in Figs. 9, 10 and 11  of ~\citet{Otarola09}, where the maximum PWV values
occur during the Summer. {\bf On the other  hand, ~\citet{Araiza11}  showed that the aerosol
optical depth  has a seasonal variation being  higher at spring (maximum) and summer than
in the rest of the year  (c.f. their Figs. 1, 2, 3  and Table 3).
These results suggest that  the double peak in the global distribution of $\psi$ is
 due to absorption variations in the atmosphere.}

 The larger value of the centres of the clear peak for July 2006 and August 2008 relative
to the same months of the other years, shown in Fig.~\ref{centers}, suggest that the  atmosphere
was more transparent. We analysed the aerosol optical thickness reported  by ~\citet{Araiza11}
(c.f. Table 4). The larger value of the centre for July 2006 is consistent with smaller
values of the  aerosol optical thickness for July 2005 and 2007 but marginally for July 2008.
 The bigger value of the centre for August 2008 is also consistent with smaller values of
the aerosol optical depth for August 2007 and marginally for August 2006 while for August 2005
there is not data available.

\section{ Statistics of clear time  }

\begin{figure}
\includegraphics[width=\columnwidth]{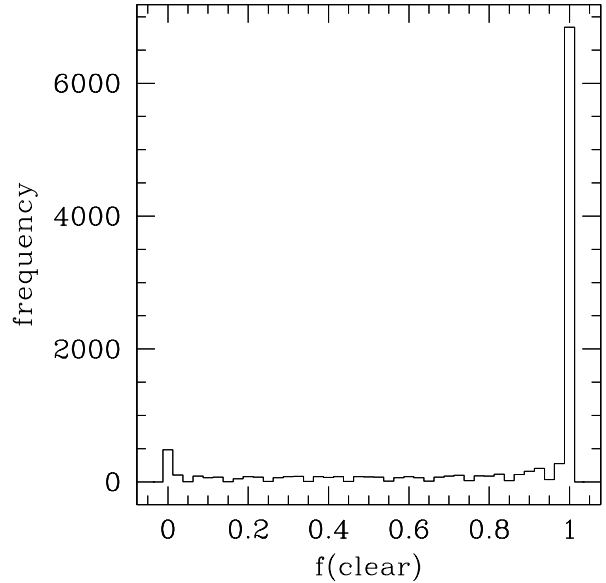}
\caption{Distribution of hourly clear fraction for the 7828 datapoints available.
\label{histograma}}
\end{figure}

  The solar radiation data observed at airmass lower than 2 is a subset of
that observed below 10. For completeness, in this analysis
we considered data with airmass less than 10.
The fraction of clear time f(clear) was  computed  for every hour of data,
accumulating 7828 h. Fig.~\ref{histograma} shows the distribution of hourly clear
fraction.  We note that it  behaves in a rather unimodal fashion: 78.6 per cent
have $f({\rm clear})=1$ while 9.5 per cent of the hours have $f({\rm clear})=0$.
The remaining fraction of data (12.5\%) have intermediate values.

\begin{figure}
\includegraphics[width=\columnwidth]{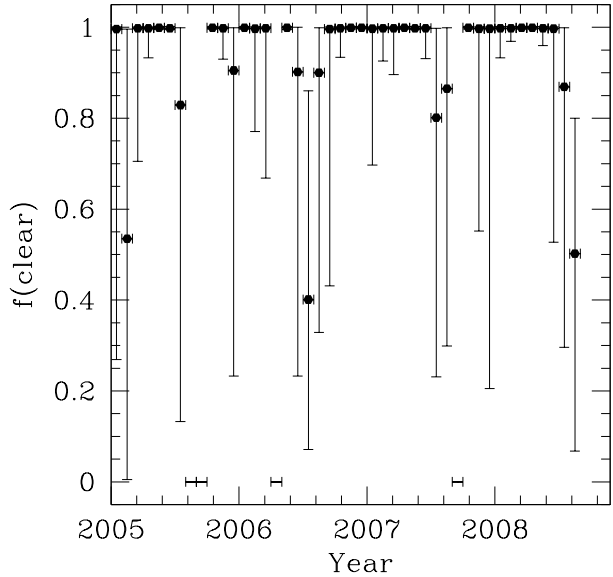}
\caption{Clear fractions for the different months. Points are
at median; bars go from 1st to 3rd quartile. The annual cycle can be appreciated.
 \label{months}}
\end{figure}

\begin{figure}
\includegraphics[width=\columnwidth]{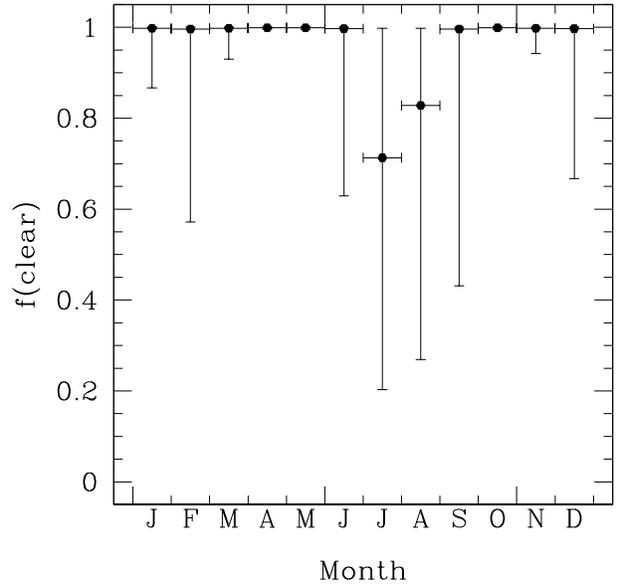}
\caption{Median and quartile values of the monthly clear fraction.}
\label {months_total}
\end{figure}

The contrast between summer and the other seasons is well illustrated in
Fig.~\ref{months}, showing the median and quartile fractions of clear time for
successive years.  The bars represent the dispersion in the data measured by the interquartile
range.   The quartiles are indicative of the  fluctuations and therefore more
representative than averages. Large  variations  are observed mainly  during
 the summer months for the whole  period.
Considerable fluctuations are also present  for 2005 in January, February and December. The latter
is not reproduced in 2006 but in 2007 there is also a large fluctuation in December.
The contrast between the spring and autumn  months, with median daily clear fractions
typically above 98 per cent,  and the cloudier  months with median clear fractions below 80 per cent is
evident. The seasonal variation can be seen with more detail in the monthly distribution of the
clear weather fraction, combining the data of different years for the same month, shown in
Fig.~\ref{months_total}. The skies are clear (f(clear)$>99$ percent) between March and
May, relatively  poor  between June and September with a minimum median value
of f(clear)$<72$ per cent) and fair between December and February  when in
the worst case 25 per cent of the time f(clear)$<57$ per cent.

\begin{figure}
\includegraphics[width=\columnwidth]{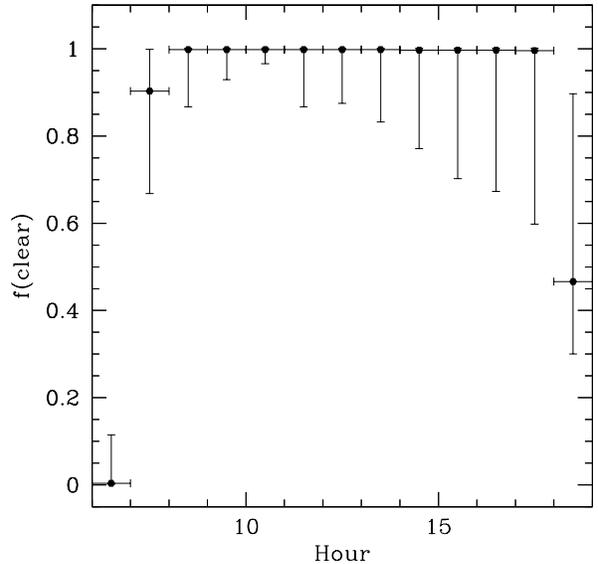}
\caption{Median and quartile values of the hourly clear fraction.
\label{hours_total}}
\end{figure}

Fig. \ref{hours_total} shows the median and quartile clear fractions as function of hour
of day.  Good conditions are more common in the mornings. The trend in our results
for daytime is consistent with that obtained by ~\citet{Erasmus02}. By analysing  the clear
fraction during day and nighttime they found that the clear fraction is highest before noon,
has a minimum  in the afternoon and increases during nighttime.  The authors associated the
afternoon  maximum in cloudiness with lifting of the inversion and cloud layer because
the site is high enough  to be located above the inversion layer at night and in the mornings.

\begin{figure}
\includegraphics[width=\columnwidth]{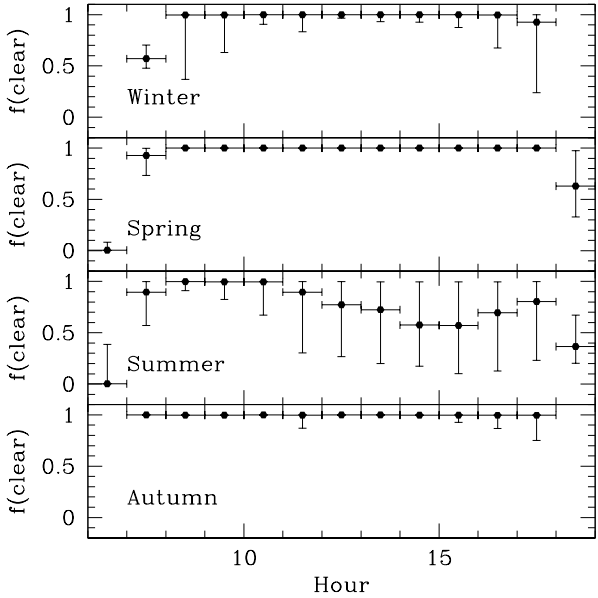}
\caption{Graph showing the median and quartile values of the fraction of clear
weather for each hour of day for each season.}
\label{hours_seasons}
\end{figure}

Fig.~\ref{hours_seasons} presents the median and quartiles of clear fraction as a
function of hour of day for the seasons subsets. Seasons were considered as follows,
winter: January, February and March; spring: Abril, May and June;
summer: July, August and September and  autumn: October, November and December.  It is
clear that during the summer the conditions are more variable than at any other
epoch of the year. In the other seasons the conditions are very stable.

\begin{figure}
\includegraphics[width=\columnwidth]{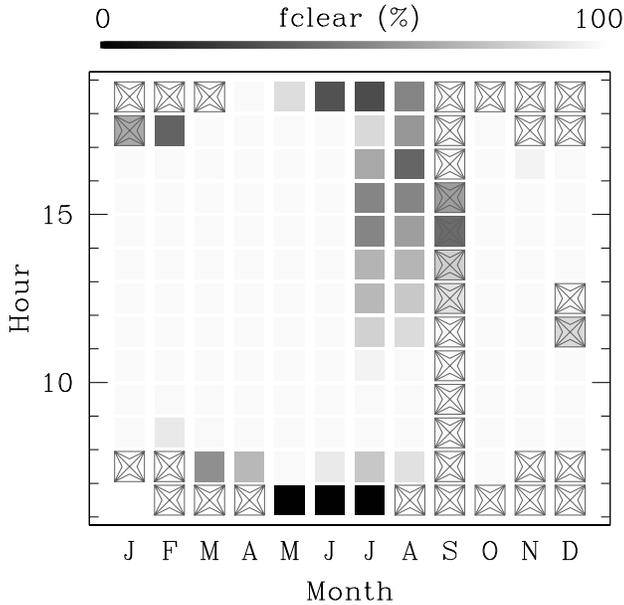}
\caption{Grey level plot showing the median fraction of clear time for
each month and hour  of day. Squares are drawn when more than 10
hour of data are available; crosses indicate less than 10 hr  of data.
\label{month_hour}}
\end{figure}

Fig.~\ref{month_hour} shows a grey level plot of the median percentage
of clear time for a given combination of month and hour of day. Squares are
drawn when more than 10 h of data are available; crosses indicate less than
10 h of data. Clear conditions  are present in the colder and drier
months, from October to June. Dark squares show cloudy weather,  clearly dominant in
the afternoons of the summer months, from July to September.

 We  repeated the analysis for airmass lower than 2. An equivalent
 histogram to that shown in Fig. \ref {histograma},  was created  by computing
the fraction of  f(clear) for every hour, adding 5211 h.  As expected, it also
has  an almost unimodal distribution: 82.5 per cent have
f(clear) = 1  while 6.7 per cent of hours  f(clear) = 0. The remaining  fraction of
data have intermediate
values.  The  values  of f(clear)  obtained for the periodicities presented
in this section  are very similar but with  less dispersion. In fact,  in the analysis
per  hour  the difference in median values are
within 0.1 per cent. For the   analysis per month  the differences are  also
in that range  except for July  and August  with differences  between
 0.3 to 13  per cent, with a  maximum of 20 per cent for July 2006.
The lower values obtained for the global distribution and for different
periods  can  be explained by the presence of clouds formed at airmass
$ 2 < z < 10$. The equivalent  grey level plot of  Fig.~\ref{month_hour}  for
 airmass less  than 2 (not shown)  does not  include the  contribution  of
clouds formed in the  early morning and late afternoon hours,
specially during  the summer months.

\section{Summary and conclusions}

Using the solar radiation data recorded by the TMT site-testing group at SPM observatory,
we applied a novel method developed by ~~\citet{Carrasco09} to estimate the time when the
sky is clear of clouds.
From the global normalized observed distribution of $\psi$: the solar flux
$F(t)$, divided by the nominal  solar flux at the top of the atmosphere $F_{\sun}\cos\theta_{\sun}(t)$
we obtained that  82.4 per cent of the time the sky is clear of clouds for airmasses $ z \le 2$.
This result is consistent
within 6 per cent  with the independent estimation of cloud cover using satellite data by ~\citet{Erasmus02}
who calculated a nighttime  useful fraction of 81 per cent and estimated that the daily clear fraction is
about 5 per cent less.

  By analyzing  the fits to the  histograms of $\psi$ per month, we  found that there is an annual
seasonal effect: the centre of the clear peak has a maximum  of   0.889 in February, reaches a minimum  of
 0.761 in July  and  increases  to  0.885 towards the end of the year. As an example we
presented the fits to the  histograms for July and November obtained using the complete
data set. The centre of the peak for July is 0.761 and for  November is 0.841 but
the amount of radiation  corresponding  to $\psi=0.761$ received at the site in July  is
277 Wm$^{-2}$ more than that in November.  This result  is consistent with the seasonal variation in
PWV measured at the site for several years  obtained  by \citet{Otarola09} and  \citet{Otarola10}.
The difference can be explained by the  presence  of more aerosols in July that absorb the incident
radiation in the atmosphere. \citet{Araiza11}  found that there are more aerosols {\bf  during  spring
 (maximum) and summer than in the rest of the year}. These results suggest that
when the sky is clear of clouds the value of $\psi_0$  is related to the  atmosphere transparency.

For completeness we carried  out a statistical analysis of the  f(clear)
 for airmass lower than 10. First, we obtained  a non symmetrical   distribution of hourly
clear time   showing that 78.6 per cent of the time the  sky is completely clear at this airmass
interval.  We calculated  the first, second and third quartile
of f(clear)  for different periodicities. We presented the results for  each month of the
four-year observing period. An annual cycle is clearly noticeable: large fluctuations are observed
mainly during the summer months. Big variations are also present for 2005 in January, February and December
but this trend is not reproduced in 2006.
The monthly distribution of f(clear) was obtained by combining the data of the same months for the
whole period. The median of f(clear) is $\sim$ 0.99 between March and May, is relatively poor between
June and September, with a minimum median value of about 0.72 in July,  and fair during
December and February. In addition, we calculated the quartiles of
f(clear) as a function of hour of day using the complete data set: good conditions are more common
in the mornings, f(clear) is highest before noon and decreases towards the afternoon. We also
carried out the same  analysis  of hour of day  for the seasons subset. It is apparent that the conditions
are very stable for all the seasons except in the summer when there is more variability. To summarise
our results we created a grey level plot, where f(clear) is represented by the grey intensity,  indicating
the median fraction of clear time for each month and hour of day: clear conditions exist in the colder
and drier months, from October to June while  cloudy weather is present in the afternoons of the summer months.

The fit to the histograms of $\psi$ developed by \citet{Carrasco09} for Sierra Negra also
reproduced the SPM data showing that this method might be generalized to other observatory
sites. Furthermore, the consistency of our results with those obtained by other authors  shows the
great potential of our method  as cloud cover is a crucial  parameter for astronomical characterization
of any site and can be estimated from in situ measurements.

\section*{Acknowledgments}
The authors acknowledge the kindness of the TMT site-testing group. The authors also thank
G. Sanders, G. Djorgovski, A. Walker and M. Sch\"ock and  for their  permission to use the results
from the ~\citet{Erasmus02}  report for SPM.
This work was partially supported by CONACyT and PAPIIT through grants number 58291 and
IN107109, respectively.

\label{lastpage}

\begin{thebibliography}{}

\bibitem[\protect\citeauthoryear{Araiza \& Cruz-Gonz\'alez}{2011}]{Araiza11}
Araiza M.R. \& Cruz-Gonz\'alez I., Rev. Mex. AA 2011, 47, 409

\bibitem[\protect\citeauthoryear{Carrasco et al.}{2009}]{Carrasco09}
Carrasco E., Carrami\~ nana A, Avila R., Guti\'errez C., Avil\'es J.L., Reyes J.
Meza J. \& Yam. O., 2009, Mon. Not. R. Astron. Soc., 398, 407

\bibitem[\protect\citeauthoryear{Cruz-Gonz\'alez et al.}{2003}]{CruzGlez03}
Cruz-Gonz\'alez I., Avila R. \&  Tapia M., eds, 2003, Rev. Mex. AA (SC), 19

\bibitem[\protect\citeauthoryear{Cruz-Gonz\'alez, et al.}{2007}]{CruzGlez07}
Cruz-Gonz\'alez I., Echevarr\'{\i}a J. \&  Hiriart D., eds, 2007, Rev. Mex. AA (SC), 31

\bibitem[\protect\citeauthoryear{Erasmus \& Van Staden}{Erasmus \& Van Staden}{2002}]{Erasmus02}
 Erasmus A, Van Staden C.~A., 2002, ``A satellite survey of cloud cover and water
vapor in the western USA and Northen Mexico. A study conducted for the CELT
project.'', internal report

\bibitem[\protect\citeauthoryear {Fr\"ohlich \& Lean} {1998}] {Frohlich98}
Fr\"ohlich, C. \& Lean, J., 1998, Geophys. Res. Let. 25, 4377

\bibitem[\protect\citeauthoryear{Hiriart et al.}{1997}]{Hiriart97}
Hiriart D. et al., 1997, Rev. Mex. AA, 33, p. 59

\bibitem[\protect\citeauthoryear{Hiriart et al.}{2003}]{Hiriart03}
Hiriart D. et al., 2003,  Rev. Mex. AA (SC), 19, 90

\bibitem[\protect\citeauthoryear{Ot\'arola et al.}{2009}]{Otarola09}
Ot\'arola A. et al., 2009,  Rev. Mex. AA, 45, 161

\bibitem[\protect\citeauthoryear{Ot\'arola et al.}{2010}]{Otarola10}
Ot\'arola A. et al., 2010, Publ. Astr. Soc. Pac., 122, 470

\bibitem[\protect\citeauthoryear{Sch\"ock et al.}{2009}]{Schock09}
Sch\"ock M.  et al., 2009,  Publ. Astr. Soc. Pac., 121, 384

\bibitem[\protect\citeauthoryear{Skidmore et al.}{2009}]{Skidmore09}
Skidmore et al. 2009, PASP 121, 1151

\bibitem[\protect\citeauthoryear{Tapia}{1992}]{Tapia92}
Tapia, M., 1992, Rev. Mex. AA 24, 179

\bibitem[\protect\citeauthoryear{Tapia}{2003}]{Tapia03}
Tapia M., 2003, Rev. Mex. AA (SC), 19, 75

\bibitem[\protect\citeauthoryear{Tapia, Hiriart \& Cruz-Gonz\'alez}{2007}]{Tapia07}
Tapia M., Hiriart D., Richer M. \& Cruz-Gonz\'alez, I. 2007,
Rev. Mex. AA (SC), 31, 47


\end{thebibliography}
\end{document}